\documentclass[12pt]{article}
\usepackage{epsf,epsfig,rotating}
\usepackage{xcolor}
\usepackage{cite}
\begin{document}

\begin{center}

\begin{large}
\textbf
{A light CP-odd Higgs boson of the MSSM Higgs sector extended by dimension-six operators}
\end{large}

\vspace{3mm}
M.~N. Dubinin, E.~Yu. Fedotova

\vspace{2mm}
{\it Skobeltsyn 
Institute of Nuclear Physics, Lomonosov Moscow State University,\\
  Leninskie Gory, 119991 Moscow, Russian Federation}

\end{center}

\begin{abstract}
The possibility of identification of an observable CMS $\mu^+ \mu^-$ excess at 28 GeV in the channel $pp\to \mu^+ \mu^- b \bar b$ at $\sqrt{s}$=8 TeV and 13 TeV as a manifestation of one of the minimal supersymmetric standard model (MSSM) Higgs bosons is investigated. The MSSM parametric scenarios in the regime of large threshold corrections involving low-mass CP-odd scalar, a 125 GeV CP-even scalar and other Higgs bosons with suitable masses are found, where the alignment limit conditions for the Higgs couplings are respected. Perturbative unitarity bounds and constraints on the electroweak vacuum stability are discussed in the regime of substantial couplings with the top- and bottom superpartners.
LHC phenomenology including top-quark decay in such a regime is analyzed.  


\end{abstract}



\section{\large INTRODUCTION}
\label{intro}

Recently the CMS experiment published results of a search for resonances decaying to a muon pair in the mass 
12--70 GeV produced in association with a $b$ quark jet and an additional jet \cite{28GeV}.
The analyses are based on data from $pp$ collisions at $\sqrt{s}=$8 TeV and 13 TeV corresponding to integrated luminosities of 19.7 and 35.9 fb$^{-1}$, respectively. Two specific event categories were analyzed, first category includes a $b$ quark jet with rapidity $|\eta|<$2.4 and at least one forward jet $|\eta|>$2.4 while second category events are required to have two jets with $|\eta|<$2.4, one of them is a $b$ quark jet and no forward jets at $|\eta|>$2.4. At $\sqrt{s}=$8 TeV an excess of $\mu^+ \mu^-$ events is observed at the dimuon invariant mass 28 GeV with local signal significances 4.2 and 2.9 standard deviations for the first and second event categories, respectively.
Reconstructed cross sections of the excesses are $4.1 \pm 1.4$ fb and $4.2\pm 1.7$ fb for these categories.
At $\sqrt{s}=$13 TeV the first event category demonstrates an excess of 2.0$\sigma$ (cross section $1.4 \pm 0.9$ fb), while the second event category results in a 1.4$\sigma$ deficit ($-1.5 \pm 1.0 $ fb).

In the following a search for the MSSM scenarios
including light CP-odd scalar $A$ with mass 28 GeV in combination with CP-even Higgs boson with mass 125 GeV which could be identified as a resonance observed by ATLAS and CMS Collaborations \cite{lhc-higgs1, lhc-higgs2} with properties consistent \cite{jhep_atlas_cms1, jhep_atlas_cms2} with predictions of the Standard Model (SM), and three heavy enough bosons of the MSSM Higgs sector (heavy CP-even boson $H$ and charged Higgs boson $H^\pm$) will be performed. 
Although the possibility of a light CP-odd Higgs boson in the commonly
considered MSSM scenarios is excluded, the situation can be changed dramatically when
higher-dimensional operators in the Higgs sector are taken into account \cite{own3}.
At present time combined results of ATLAS and CMS Collaborations at $\sqrt{s}=$7 and 8 TeV \cite{jhep_atlas_cms1, jhep_atlas_cms2}  
and recent results at $\sqrt{s}=13$ TeV \cite{LC_atlas_cms1, LC_atlas_cms2} 
for the Higgs boson production cross sections and decay rates still leave a room for meaningful contributions of physics beyond the SM. 
The MSSM mass spectrum of the Higgs sector which is defined at the tree-level  by the parameters $m_{A}$ (the CP-odd Higgs boson mass) and $\tan \beta=v_2/v_1$ (the ratio of vev's in the Higgs isodoublets) \cite{mssm1, mssm2},  is strongly influenced by radiative corrections coming (in the natural MSSM scenarios) from the side of the third generation SM fermions (top- and bottom-quarks) and scalar quarks.
Mass spectrum of scalars must respect both the theoretical and experimental constraints, which are perturbative unitarity, vacuum stability, alignment limit for the scalar couplings from the theoretical side and
model-independent experimental limits for the ${H^\pm}$ and $H$ masses \cite{pdg} which are 
$m_{H^\pm}>80$ GeV,
$m_H>92.8$ GeV.
Most recent model-dependent constraints from radiative $B$-meson decays give stronger limits for $m_{H^\pm}$ \cite{B2017}.
Searches for a light pseudoscalar have been performed recently also in final states with $\tau^+ \tau^-$ produced in association
with a bottom quark \cite{searchtau19} (see also  \cite{searchtau17}) and in final states with a single photon and missing energy \cite{belle19}. Corresponding upper limits in $\tau^+ \tau^-$ invariant mass range of 25--70 GeV at 95\%
confidence level are from 250 pb to 44 pb \cite{searchtau19}.

A significant interest in dimuon resonance at 28 GeV was shown in a recent study of the muon anomalous magnetic moment \cite{vysotsky}, where a new resonance could explain the deviation of the measured $(g-2)_\mu$ from the SM value. A recent detailed study of possible searches for light Higgs bosons at the high luminosity LHC in the framework of the next-to minimal supersymmetric standard model (NMSSM) can be found in \cite{kazakov}. Nine-dimensional parameter space of the semi-constrained NMSSM provides greater freedom of choice in comparison with five-dimensional MSSM parameter space studied below, when the only candidate to explain the dimuon excess is the CP-odd Higgs boson.

In this paper the possibilities of the existence of the low mass pseudoscalar in the two-doublet MSSM Higgs sector are discussed. 
The paper is organized as follows.
In Section \ref{ex} some examples of the effective couplings structure in the effective Higgs potential are given. 
Perturbative unitarity and vacuum stability constraints are discussed in Section \ref{pert}.
Numerical results in the frameworks of several MSSM parametric scenarios (the so-called benchmark scenarios \cite{benchmark}) evaluated for a number of benchmark points are discussed in Section \ref{num}.
We provide a summary of these results in Section \ref{concl}.

\section{\large EFFECTIVE FIELD THEORY FRAMEWORK}
\label{eft}
\subsection{Effective couplings structure} 
\label{ex} 
Calculations of the two-doublet potential were performed using the effective potential method. The effective Higgs potential in the Coleman-Weinberg framework \cite{cw73} 
can be presented as a sum of terms of all orders of perturbation theory 
\begin{equation}
U({\rm 1-loop})=U^{(2)}+U^{(4)}+U^{(6)}+\ldots,
\label{U}
\end{equation}
where 
\begin{eqnarray}
U^{(2)} &=&
- \, \mu_1^2 (\Phi_1^\dagger\Phi_1) - \, \mu_2^2 (\Phi_2^\dagger
\Phi_2) - [ \mu_{12}^2 (\Phi_1^\dagger \Phi_2) +h.c.], \\
U^{(4)} &=& \lambda_1
(\Phi_1^\dagger \Phi_1)^2
      +\lambda_2 (\Phi_2^\dagger \Phi_2)^2
+ \lambda_3 (\Phi_1^\dagger \Phi_1)(\Phi_2^\dagger \Phi_2)
+
\lambda_4 (\Phi_1^\dagger \Phi_2)(\Phi_2^\dagger \Phi_1) \nonumber \\
&+& [\lambda_5/2
       (\Phi_1^\dagger \Phi_2)(\Phi_1^\dagger\Phi_2)  
  +\lambda_6
(\Phi^\dagger_1 \Phi_1)(\Phi^\dagger_1 \Phi_2)+\lambda_7 (\Phi^\dagger_2 \Phi_2)(\Phi^\dagger_1 \Phi_2)+h.c.],  \label{U4}\\
U^{(6)} &=& \kappa_1 (\Phi^\dagger_1 \Phi_1)^3
+\kappa_2 (\Phi^\dagger_2 \Phi_2)^3+
\kappa_3 (\Phi^\dagger_1 \Phi_1)^2 (\Phi^\dagger_2 \Phi_2)  
+\kappa_4 (\Phi^\dagger_1 \Phi_1) (\Phi^\dagger_2 \Phi_2)^2\nonumber \\
&+&\kappa_5 (\Phi^\dagger_1 \Phi_1) (\Phi^\dagger_1 \Phi_2) (\Phi^\dagger_2 \Phi_1)  
+\kappa_6 (\Phi^\dagger_1 \Phi_2) (\Phi^\dagger_2 \Phi_1) (\Phi^\dagger_2 \Phi_2) \nonumber \\
&+& [\kappa_7 (\Phi^\dagger_1 \Phi_2)^3  
+\kappa_8 (\Phi^\dagger_1 \Phi_1)^2 (\Phi^\dagger_1 \Phi_2)+\kappa_9 (\Phi^\dagger_1 \Phi_1) (\Phi^\dagger_1 \Phi_2)^2 \nonumber \\
&+&\kappa_{10} (\Phi^\dagger_1 \Phi_2)^2 (\Phi^\dagger_2 \Phi_2) 
+\kappa_{11} (\Phi^\dagger_1 \Phi_2)^2 (\Phi^\dagger_2 \Phi_1)  
+\kappa_{12} (\Phi^\dagger_1 \Phi_2) (\Phi^\dagger_2 \Phi_2)^2 \nonumber \\
&+&\kappa_{13} (\Phi^\dagger_1 \Phi_1) (\Phi^\dagger_1 \Phi_2) (\Phi^\dagger_2 \Phi_2)+h.c.],\label{U6}
\end{eqnarray}
and
\begin{equation}
\label{dublets} \Phi_i= \left(\begin{array}{c} \phi_i^+(x) \\ \phi_i^0(x) \end{array} \right)=\left(\begin{array}{c} -i \omega_i^+ \\ \frac{1}{\sqrt{2}} (v_i+\eta_i+i \chi_i) \end{array} \right), \qquad
i=1,2
\end{equation}
are Higgs doublets with the $SU(2)$ field states and $v_1=v \cos \beta$, $v_2=v \sin \beta$ ($v=\sqrt{v_1^2+v_2^2}=246$ GeV) are vacuum expectation values of them.

The parametrization of the MSSM soft supersymmetry breaking sector most common in the MSSM benchmark scenarios \cite{benchmark} uses five-dimensional MSSM parameter space $(m_A,\tan \beta, M_S, A_{t,b}, \mu)$, where $M_S$ is the quark superpartners mass scale, $A_{t,b}$ are the trilinear soft supersymmetry breaking parameters and $\mu$ is the Higgs superfield mass parameter. The dimension-six operators may play an important role if $A_{t,b}, \mu$ satisfy the following conditions
\begin{equation}
\begin{tabular}{ccc}
$|\mu| m_{top} \cot \beta \approx M_S^2$, & $|A_t| m_{top} \approx M_S^2$, & $|\mu A_t| m_{top}^2 \cot \beta \approx M_S^4$, \\
$|\mu| m_b \tan \beta \approx M_S^2$, & $|A_b| m_b \approx M_S^2$, & $|\mu A_b| m_b^2 \tan \beta \approx M_S^4$, \label{6_dim_cond} 
\end{tabular} 
\end{equation}
i.e. $A_{t,b}$ and $\mu$ range is of the order of a few TeV or more in combination with moderate $M_S$  at the TeV scale. 
Such a situation is rather unusual in most of the MSSM scenarios. 
Radiative corrections to parameters $\lambda_i$, $i$=1,...7 in the effective field theory framework have been analyzed in Refs. \cite{haber_hempfling, eff_potential1, eff_potential2, eff_potential3, eff_potential4, own1, own2, sim_sc}. Radiative corrections to the parameters $\kappa_i$, $i=$1,...13 in the approximation of degenerate squark masses have been obtained in Ref. \cite{dim_six}.
An example of the one-loop RG-improved threshold correction structure for $\lambda_1$ and the threshold correction for $\kappa_1$ in the form which uses $A_{t,b,}/M_S$ and $\mu/M_S$ power terms is
\begin{eqnarray}
-\frac{\Delta \lambda_1^{\rm thr}}{2} &=& \frac{3}{32\pi^2} \Big[ h^4_b \frac{|A_b|^2}{M^2_{S}}\left(2-\frac{|A_b|^2}{6 M^{\,2}_{S}}\right)
-h^4_t\frac{|\,\mu|^4}{6 M^{\,4}_{S}} 
+ \,2 h_b^4 l +\,\frac{g_2^2+g_1^2}{4\,M^{\,2}_{S}} \nonumber \\
&\times&(h^2_t|\,\mu|^2-h^2_b|A_b|^2) \Big]  
+ \, \frac{1}{768\pi^2}\,
\left(11 g_1^4 + 9g_2^4 - 36 \,(g_1^2+g_2^2)\,h_b^2\right) l ,  \\
\Delta \kappa_1^{\rm thr} &=& \frac{h_b^6}{32 M_S^2 \pi^2} \left(2-\frac{3 |A_b|^2}{M_S^2}+\frac{|A_b|^4}{M_S^4}-\frac{|A_b|^6}{10 M_S^6} \right) \nonumber \\
&-& h_b^4 \frac{g_1^2+g_2^2}{128 M_S^2 \pi^2} \left( 3-3\frac{|A_b|^2}{M_S^2}+\frac{|A_b|^4}{2 M_S^4} \right)  
+ \frac{h_b^2}{512 M_S^2 \pi^2} \nonumber \\ 
&\times& \left(\frac{5}{3}g_1^4+2 g_1^2 g_2^2+3g_2^4 \right) 
 \left(1-\frac{|A_b|^2}{2 M_S^2} \right)
-h_t^6 \frac{|\mu|^6}{320 M_S^8 \pi^2} 
+ h_t^4 \frac{(g_1^2+g_2^2)|\mu|^4}{256 M_S^6 \pi^2} \nonumber\\
&-&h_t^2 \frac{(17 g_1^4-6g_1^2 g_2^2+9g_2^4) |\mu|^2}{3072 M_S^4 \pi^2} + \frac{g_1^2}{1024 M_S^2 \pi^2} (g_1^4-g_2^4),
\end{eqnarray}
where $l\equiv\ln\left(\frac{ M^{\,2}_{S}}{\sigma^2}\right)$, $\sigma=m_{top}$ is the renormalization scale, $h_{t}=\frac{g_2 m_{top}}{\sqrt{2}m_W \sin \beta}$ and $h_{b}=\frac{g_2 m_{b}}{\sqrt{2}m_W \cos \beta}$ are the Yukawa couplings.
 One can notice inspecting such explicit forms that radiative corrections $\Delta \kappa^{\rm thr}$ begin to play a significant role if the conditions (\ref{6_dim_cond}) are true.

All Higgs masses acquire additional contributions coming from $\lambda_i$ and $\kappa_i$ (see \cite{dim_six}). For example, 
the charged Higgs boson mass is shifted from the tree-level value by $\Delta \lambda_{4,5}$ and $\kappa_i$, $i$=5,6,7,9,10,11
\begin{eqnarray}
m_{H^\pm}^2&=&m_W^2+m_A^2-\frac{v^2}{2} ({\rm Re}\Delta \lambda_5-\Delta \lambda_4) 
+ \frac{v^4}{4} [c_\beta^2(2 {\rm Re}\kappa_9-\kappa_5)\nonumber \\
&+&s_\beta^2(2 {\rm Re}\kappa_{10}-\kappa_6)-s_{2 \beta} ({\rm Re}\kappa_{11}-3{\rm Re}\kappa_7)] . 
\label{mcharged}
\end{eqnarray}

\subsection{Perturbative unitarity and vacuum stability}

\label{pert}

Rather unusual $A_{t,b}$ and $\mu$ parameters range of a few TeV simultaneously raises questions about perturbative unitarity of the model and vacuum stability in such a regime. A discussion of the perturbative unitarity constraints \cite{unitarity_1} in the two-Higgs doublet model can be found, for example, in Refs. \cite{unitarity_21, unitarity_22}. For a single 2$\to$2 scattering amplitude they are imposed by an analysis in terms of the partial wave decomposition
\begin{equation}
M(s) = 16\pi \sum_{l=0}^{\infty} (2l+1) P_l(\cos \theta) a_l(s),
\label{partial}
\end{equation}
where $s$ is a Mandelstam variable, $P_l$ are Legendre polynomials and $a_l$ is the partial amplitude. From the differential cross section formula in the massless limit $d\sigma/d\Omega = |M|^2/64\pi^2 s$, it follows that
\begin{equation}
\sigma = \frac{16\pi}{s} \sum_{l=0}^\infty (2l+1) |a_l(s)|^2,
\end{equation}
so the optical theorem for $\sigma$ gives the unitarity constraint for the real part of the amplitude
\begin{equation}
Re(a_l)^2 + Im(a_l)^2 = |a_l|^2 = Im(a_l) \quad \Longleftrightarrow \quad |Re(a_l)| \leq \frac{1}{2}
\end{equation} 
at tree level.
This condition must be satisfied by any eigenvalue $x_i$ of the scattering matrix derived by including all possible combinations of scalars in the initial and final states. The general structure of a matrix form for the MSSM can be found in \cite{staub}. General unitarity calculations for the case of the two-Higgs doublet model (THDM) are performed 
using the inverse expression of (\ref{partial})
\begin{equation}
a_l(s) = \frac{1}{32\pi} \int_{-1}^{1} d\cos \theta P_l(\cos \theta) M(s)
\end{equation}
for the $s$-wave amplitude $a_0(s)$. In the four-scalar channel $H_1 H_2 \to H_3 H_4$ in the massless limit one can get $a_0(s) < V_4(H_1 H_2 H_3 H_4)/16\pi$ 
in the limit $s \rightarrow \infty$\footnote{Such approximations are oftenly not suitable, see \cite{staub} and refs therein.},
where $V_4$ is the quartic vertex, $t,u$-channel diagrams with the triple vertices of scalars $V_3(H_1 H_2 H_3)$ are omitted. It follows that an approximate unitarity constraint for the quartic coupling is $V_4(H_1 H_2 H_3 H_4) < 8\pi$. In the general case of numerous quartic contributions to 2$\to$2 channels, a tree-level scattering matrix for scalars is constructed in the mass basis and diagonalized imposing then constraints on the eigenvalues \cite{unitarity_1}. This procedure can be performed \cite{unitarity_21, unitarity_22} in the THDM basis of $SU(2)$ states 
if it is related to the mass basis by a unitary transformation.
Computation of the
two-body elastic scatterings of Nambu-Goldstone bosons and physical Higgs bosons for the general THDM and 
the analytic formulae for the block-diagonalized S-wave amplitude matrix

\begin{equation}
\label{a00}
a_0(s)=\frac{1}{16 \pi} {\rm Diag}
(
X_{4\times4}, Y_{4\times4}, Z_{3\times3}, Z_{3\times3}
)
\end{equation}
can be found in \cite{kanemura}. The eigenvalues of submatrices $X_{4\times4}, Y_{4\times4}, Z_{3\times3}$ 
must respect the unitarity bound
\begin{equation}
|{\rm Re}(x_i)|< 1
\label{pertb}
\end{equation}
where
\begin{equation}
\label{eigenv}
x_i=\{{\rm Eigenvalues}(X_{4\times4}), {\rm Eigenvalues}(Y_{4\times4}), {\rm Eigenvalues}(Z_{3\times3}), \lambda_3-\lambda_4\}.
\end{equation}
In the general case, analytic expressions for the eigenvalues $x_i$  
are very complicated, so the following analysis will be performed numerically.
Note that perturbative unitarity constraints for the one-loop effective potential of the MSSM decomposed to the dimension-four effective operators \cite{staub} are weaker than the vacuum stability bounds. 

Recent analyses of the vacuum stability of the MSSM electroweak (EW) vacuum carried out with non-zero vacuum expectation values of SUSY fields, performed by means of the polynomial homotopy continuation method \cite{staub, vacuum}, uses the stability criterion which is realized if the deepest minimum of the effective potential
$
V=V_F+V_D+V_{soft}
$
coincides with the EW minimum (here 
$V_F$ comes from the superpotential, 
$V_D$ comes from the gauge structure of the model,
 $V_{soft}$ includes soft SUSY breaking terms).
An approximate bound used to judge whether a parameter point might be sufficiently long-lived (a 'heuristic' bound \cite{vacuum}) is
\begin{equation}
\frac{{\rm max}(A_{t,b}, \mu)}{{\rm min}(m_{Q_3,U_3})} \leq 3.
\label{heuristic}
\end{equation}
This bound is obtained for the dimension-four potential terms. The case of electroweak minimumum of the Higgs potential decomposed to dimension-six effective operators was discussed in \cite{own-min}.

\section{\large NUMERICAL ANALYSIS}

\label{num}

The numerical analysis is performed in a framework of the EFT approach with radiative corrections to the Higgs sector calculated in Refs. \cite{own1, own2, dim_six},
where all non-SM particles share a common mass $M_S$ and the effective theory below this scale is the THDM.
Using the five-dimensional MSSM parameter space practically identical to the parameter space of the natural MSSM benchmark scenarios \cite{benchmark}, radiative corrections to $\lambda_i$ $i$=1,...7 and $\kappa_j$, $j$=1,...13, are evaluated using the set $(m_A,\tan \beta, M_S, A_{t,b}, \mu)$. This scan of the MSSM parameter space allows one to find a number of benchmark points which respect the following requirements:

(A) {\it The mass basis for the Higgs bosons exists.} In other words, after diagonalization squared masses of the scalars are positively defined. A detailed analysis can be found in \cite{own3}. In the EFT framework masses of Higgs bosons are evaluated as eigenvalues of mass matrices containing all factors in front of the  lagrangian terms of dimension two in terms of $SU(2)$ fields. In some regions of the MSSM parameter space, negative eigenvalues of mass matrices can appear so the mass basis of physical scalars may not exist. Note that masses of CP-even bosons $h$ and $H$ do not depend on mixing angle $\alpha$ in the diagonalization procedure that we are using.

(B) {\it Yukawa couplings respect the alignment limit} \cite{alignment1, alignment2}. In other words, observable couplings of the Higgs boson with mass 125 GeV are SM-like, i.e. $g_{\Phi uu} \approx g_{\Phi dd} \approx g_{\Phi VV} \approx 1$, where $\Phi$ is $h$ or $H$. The couplings of $\Phi$ with SM quarks and gauge bosons are presented below. 
One can see that $h$-alignment limit is
realized if $\beta-\alpha \approx \pi/2$, while the $H$-alignment limit is valid if $\beta \approx \alpha$. The condition for $h$- and $H$-alignment limits valid simultaneously is $\tan 2\alpha \approx \tan 2 \beta$. 
\begin{table}[h!]
\begin{center}
\label{tab:gc}
\begin{tabular}{cccc}
\hline
\hline
 & $g_{\Phi uu}$ & $g_{\Phi dd}$ & $g_{\Phi VV}$ \\
\hline
$h$ & $\cos \alpha/\sin \beta$ & $-\sin \alpha/ \cos \beta$ & $\sin(\beta - \alpha)$\\ 
$H$ & $\sin \alpha/\sin \beta$ & $\cos \alpha/
\cos \beta$ & $\cos (\beta - \alpha)$\\
\hline
\hline
\end{tabular}
\end{center}
\end{table}
Note here that mixing angle $\alpha$ is an input parameter of the model specified
by equation
\begin{equation}
\tan 2 \alpha=\frac{(m_Z^2+m_A^2)s_{2 \beta}-2 \Delta {\cal M}_{12}^2}{(m_A^2-m_Z^2)c_{2 \beta}-\Delta{\cal M}_{11}^2+\Delta{\cal M}_{22}^2}
\label{alpha}
\end{equation}
and is restricted by region $-\pi/2 < \alpha < 0$. Here $\Delta {\cal M}_{ij}$ denote radiative corrections to the mass matrix, $m_{A}$ $(m_Z)$ is the mass of  pseudoscalar boson ($Z$-boson).

(C) $S$-wave partial amplitudes for the quartic couplings of Higgs bosons are restricted from above {\it respecting the perturbative unitarity constraint}.

(D) {\it Electroweak vacuum stability is respected}.

\subsection{Selection of model parameters}

Selection of model parameters that respect the requirements (A) and (B) was made as follows: when the CP-odd scalar mass is fixed at $m_A=$28 GeV, for some ($M_S$, $\tan \beta$) set taken fixed we scan ($A_{t,b}$, $\mu$) parameter space to find benchmark points (BP's) when the mass of either $h$ or $H$ is equal to 125 GeV in the alignment limit.
Parameter sets for basic BP's which we are using in the following cross section calculations are shown in Table \ref{tab:podbor}. 
It is possible to find such ($A_{t,b}$, $\mu$) sets at $M_S$ of about 2 TeV and $\tan \beta$ around 2--5 for $m_h=$125 GeV only (i.e. only $h$-alignment limit exists).
The corresponding parameter sets (selected benchmark points) are shown in Table \ref{tab:BPs}.
The remaining $H$ and $H^\pm$ bosons are not decoupled having masses of around 130 -- 150 GeV. 
It is important to note that for $M_S$ of the order of 2 TeV and small $\tan \beta \sim$ 2--5 it is possible to select suitable parametric sets only for the case of Higgs potential decomposition up to the dimension-six effective operators. In this case $A_{t,b}, \mu$ respect Eq.(\ref{6_dim_cond}), see also Table \ref{tab:BPs}. 

\begin{table}
\caption{Illustration of a scan in the $\mu, A$ plane for several ($\tan \beta$ $M_S$) sets, $\mu, A$ variation range from zero to 10 TeV. Four requirements are checked, (1) CP-even Higgs boson mass $m_h=$125 GeV with other masses of scalars positively defined; (2) CP-even Higgs boson mass $m_H=$125 GeV with other masses of scalars positively defined; (3) $h$-alignment limit for Yukawa and gauge boson couplings is possible; (4) $H$-alignment limit is possible. Model parameter values $m_A=$28 GeV, 
$\sin \vartheta_W=$0.472, $m_Z=$91.187 GeV, $m_{top}=$173.3 GeV, $m_b=$4.92 GeV, $g_s=$1.2772.}
\label{tab:podbor}
\begin{center}
\begin{tabular}{cccccccc}
\hline
\hline
 & & \multicolumn{5}{c}{ $M_S$ (GeV)} \\ 
$\tan \beta$ & dim & & 600 & 1000 & 2000 & 3500 & 5000 \\
\hline
1 & four  & & +,+,-,- & +,+,-,- & +,-,-,- & +,-,-,- & -,-,-,-\\
  & six  & & -,-,-,- & +,+,-,- & +,+,-,- & -,-,-,- & -,-,-,-\\
\\
2 & four & & +,-,-,- & +,-,-,- & +,-,-,- & -,-,-,- & -,-,-,-\\
 & six & & +,+,-,- & +,+,-,- & \textbf{+,+,+,-} & -,+,-,- & -,+,-,-\\
\\
3 & four & &  +,-,-,- & +,-,-,- & +,-,-,- & -,-,-,- & -,-,-,-\\
& six & & +,+,-,- & +,+,-,- & \textbf{+,+,+,-} & -,+,-,- & -,-,-,-\\
\\
5 & four & & +,-,-,- & +,-,-,- & +,-,-,- & -,-,-,- & -,-,-,-\\
& six & & +,+,-,- &\textbf{+,+,+,-} & \textbf{+,+,+,-} & -,+,-,- & -,-,-,-\\
\\
15 & four & & +,-,-,- & +,-,-,- & +,-,-,- & +,-,-,- & -,-,-,-\\
& six & & +,-,-,- & +,-,-,- & +,-,-,- & +,-,-,- & -,-,-,-\\
\\
20 & four & & +,-,-,- & +,-,-,- & +,-,-,- & +,-,-,- & -,-,-,- \\
& six & & +,-,-,- & +,-,-,- & +,-,-,- & +,-,-,- & -,-,-,-  \\
\hline
\hline
\end{tabular}
\end{center}
\end{table}

\begin{table}
\begin{center}
\caption{Benchmark points (BP's) used for cross-section calculations. $m_A$=28 GeV, $m_h$=125 GeV in the alignment limit.}
\label{tab:BPs}
\begin{tabular}{ccccc}
\hline
\hline
BP & $\tan \beta$ & $M_S$ (GeV) & $A_{t,b}$ (GeV) & $\mu$ (GeV)  \\ 
\hline
1 & 2 & 2000 & 8800 & 5320  \\
2 & 3 & 2000 & 7820 & 6450 \\
3 & 5 & 1000 & 3385 & 5040 \\
4 & 5 & 2000 & 6690 & 7960 \\
\hline
\hline
\end{tabular}
\end{center}
\end{table}

The masses of five Higgs bosons, couplings, the eigenvalues given by Eq.(\ref{eigenv}) and the check of EW vacuum existence are presented in Table \ref{tab:pred}. 
One can see that while the EW vacuum exists, the condition (\ref{heuristic}) is near the limit of execution. Satisfactory fulfillment of the perturbative unitarity conditions can be taken into account keeping in mind a number of approximations made at their derivation.

\begin{table}
\begin{center}
\caption{ Masses of Higgs bosons, couplings, eigenvalues, see (\ref{eigenv}) and EW vacuum existence for BP's in Table \ref{tab:BPs}, $m_A$=28 GeV, $m_h$=125 GeV.}
\label{tab:pred}
\begin{tabular}{cccccccc}
\hline
\hline
BPs & $m_H$ (GeV) & $m_{H^\pm}$ (GeV) & $g_{\Phi uu}$ & $g_{\Phi dd}$ & $g_{\Phi VV}$ & max$|x_i|$ & EW vac \\ 
\hline
1 & 134.4 & 129.7 & 1.1 & 0.7 & 1.0 & 2.1 & +\\
2 & 132.3 & 130.0 & 1.0 & 0.9 & 1.0 & 1.6 & +\\
3 & 127.7 & 127.3 & 1.0 & 1.0 & 1.0 & 6.6 & +\\
4 & 130.4 & 131.3 & 1.0 & 1.0 & 1.0 & 1.9 & +\\
\hline
\hline
\end{tabular}
\end{center}
\end{table}

\subsection{Cross section calculations}

As already mentioned in the Introduction, in the CMS study of a new signal \cite{28GeV} at the invariant mass of 28 GeV two event categories were analyzed, with forward jet (SR1 event category) and without forward jet (SR2 event category), see details of event selection in \cite{28GeV}, Table 1. So two sets of kinematical cuts corresponding to these categories are used in the following evaluations.

Basic set corresponding to the CMS selection cuts is

\begin{eqnarray}
{\rm Muons}:&& p_T>25\; {\rm GeV}, \qquad |\eta|<2.1,\qquad m_{\mu^+ \mu^-}>12 \; {\rm GeV}, \nonumber\\
b:&& p_T>30 \; {\rm GeV}, \qquad |\eta| \leq 2.4, \nonumber \\
\bar{b}:&& p_T>30 \; {\rm GeV},\qquad 2.4 \leq |\eta| \leq 4.7 \quad ({\rm SR1}),\qquad |\eta| \leq 2.4 \quad ({\rm SR2}). \nonumber
\end{eqnarray}
In order to understand the yield of irreducible background diagrams and the sensitivity of signal separation, the following phase space cuts for the signal and signal+background diagrams of the process $pp \to \mu^+ \mu^- b \bar{b}$ were imposed:

\begin{enumerate}

\item[Cut-A:]

all irreducible background diagrams with intermediate photons and gauge bosons are omitted, phase space cuts are imposed on $b, \bar{b}$ for SR1 event category.

\item[Cut-B:]
all irreducible background diagrams with intermediate photons and gauge bosons are omitted, phase space cuts are imposed on $b, \bar{b}$ for SR2 event category, 25 GeV$ \leq m_{\mu^+ \mu^-} \leq$ 32 GeV.

\item[Cut-D:] 
complete tree level set of diagrams is calculated, phase space cuts are imposed 
for SR2 event category.
\item[Cut-E:] 
complete tree level set of diagrams is calculated, phase space cuts are imposed
for SR2 event category, 
25 GeV $ \leq m_{\mu^+ \mu^-} \leq$ 32 GeV.

\end{enumerate}

\vskip 8mm
 
\begin{figure}[h]
\center{
\includegraphics[width=0.35\linewidth]{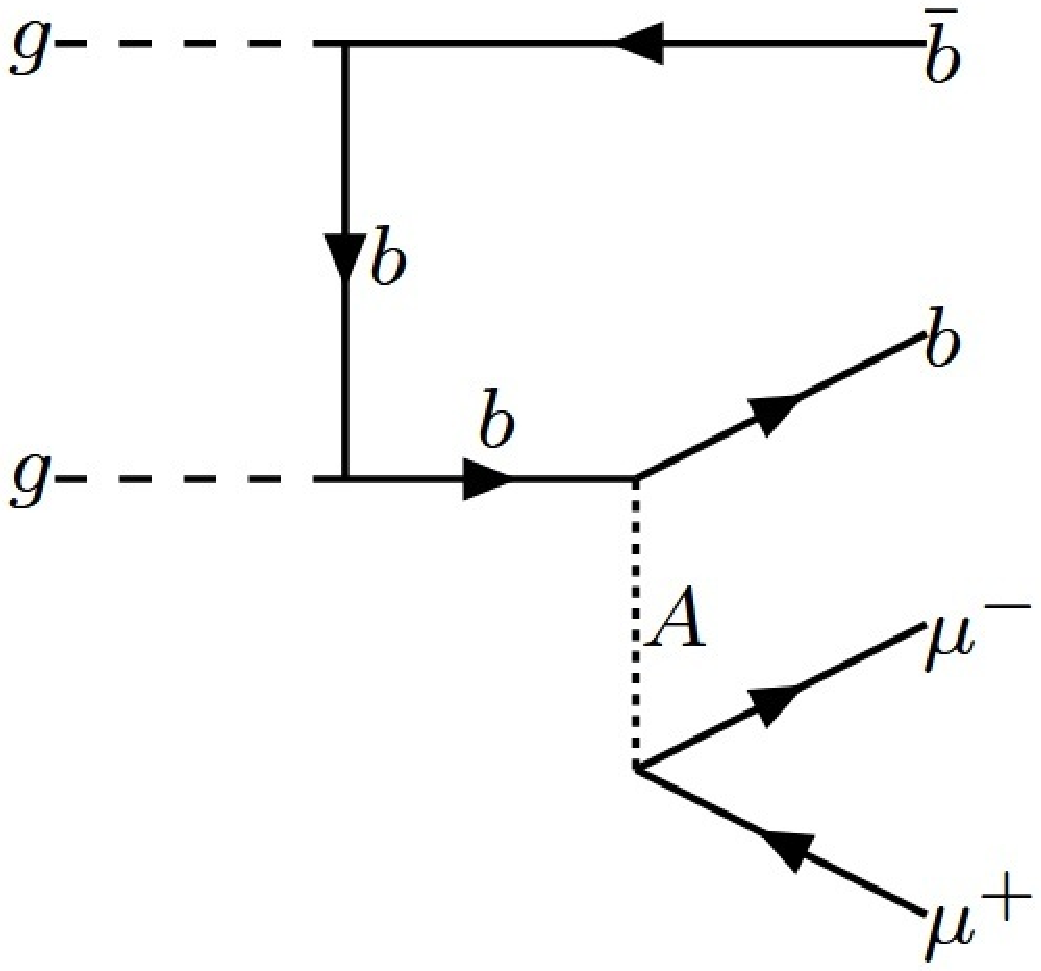} \qquad
\includegraphics[width=0.35\linewidth]{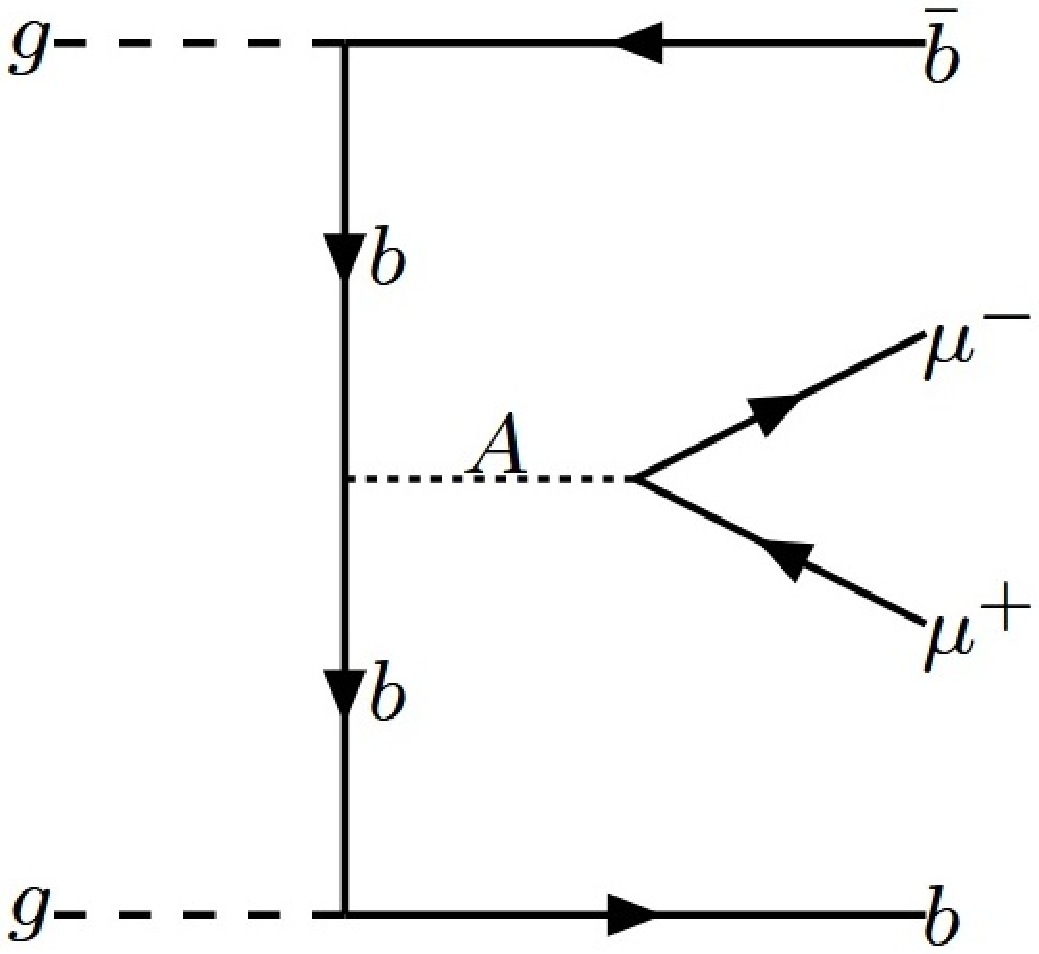}
}
\caption{Signal diagrams, where $A$ stands for the CP-odd Higgs boson field.}
\end{figure}

Complete tree-level calculations for the process $pp \to \mu^+ \mu^- b \bar{b}$ (13 partonic subprocesses) made by means of {\tt CompHEP} 
package \cite{comphep}, where the MSSM model is modified by adding dimension-six operators, demonstrated that the main contribution to the signal is given by gluon-gluon subprocess $gg \to b \bar{b} A$ with the following decay $A \to \mu^+ \mu^-$, see diagrams in Fig. 1.
Results are presented in Tables \ref{tab:sB} and \ref{tab:cs}. Subprocesses with quarks in the initial state insignificantly increase the signal cross section by 3--4\%. 'Cut-A' and 'Cut-B' 2$\to$4 process cross sections of the order of 0.01--0.40 fb at $\sqrt{s}=8$ TeV and 0.03--0.90 fb at$\sqrt{s}=$13 TeV, where practically signal only diagrams contribute, coincide well with the infinitely small width approximation $pp \to b \bar b A \times BR(A \to \mu^+ \mu^-)$ presented for corresponding cuts in Table \ref{tab:sB}. 'Cut-C', 'Cut-D' and 'Cut-E' cross sections for complete tree level set of 2$\to$4 diagrams with $A$, photon and $Z$-boson intermediate states include large flat background on which there is a small signal peak. They are sensitive to the invariant mass cut for $m_{\mu^+ \mu^-}$ when mostly background diagrams contribute, as one can see by comparing results for 'Cut-D' and 'Cut-E'.

\begin{table}
\begin{center}
\caption{$\sigma(gg \rightarrow b \bar{b} A) \times BR(A \rightarrow \mu^+ \mu^-)$ (fb), where BR$(A \rightarrow \mu^+ \mu^-)=1.6 \cdot 10^{-4}$. Cut-A and Cut-B are imposed on $b, \bar{b}$ jets for SR1 and SR2 event categories, correspondingly.
}
\label{tab:sB}
\begin{tabular}{ccccccc}
\hline
\hline
& & \multicolumn{2}{c}{SR1} & \multicolumn{2}{c}{ SR2}\\
$\sqrt{s}$ & $\tan \beta$ & $\sigma(gg \rightarrow bbA)$ (fb) & $\sigma \times BR$ (fb) & & $\sigma(gg \rightarrow bbA)$ (fb) & $\sigma \times BR$ (fb) \\
\hline
&2 & 56.63 & 0.009 & & 386.27 &  0.062  \\
8 TeV &3 & 127.19 & 0.020 & & 870.73 & 0.139\\
& 5 & 355.90 & 0.057  & & 2423.10 & 0.388\\
\\
& 2 & 165.68 & 0.026 & & 904.65  & 0.145 \\
13 TeV & 3 & 370.38 & 0.059 & & 2021.10 & 0.323\\
& 5 & 1040.88 & 0.167 & & 5640.90  & 0.903 \\
\hline
\hline
\end{tabular}
\end{center}
\end{table}

\begin{table}
\begin{center}
\caption{$\sigma(gg \rightarrow \mu^+ \mu^- b \bar{b})$ (fb) for SR1 and SR2 categories. 
}
\label{tab:cs}
\begin{tabular}{cccccccc}
\hline
\hline
& & \multicolumn{2}{c}{SR1} & &\multicolumn{3}{c}{ SR2}\\
$\sqrt{s}$ & BP & Cut-A & Cut-C & & Cut-B & Cut-D & Cut-E\\
\hline
&1 & 0.009 & 10.094 & & 0.065 & 267.240 & 0.730\\
8 TeV &2 & 0.020 & 13.242 & & 0.134 & 236.750 & 0.742\\
&3 & 0.056 & 8.814 & & 0.384 & 270.810 & 0.758\\
&4 & 0.057 & 9.800 & & 0.387 & 223.870 & 0.769\\
\\
 & 1 & 0.027 & 55.994  & & 0.148 & 571.790 & 1.887\\
 13 TeV & 2 & 0.058 & 48.692 & & 0.310 &  609.650 & 1.903\\
 & 3 & 0.165 & 53.642 & & 0.902 & 610.500 
  & 1.972\\
 & 4 & 0.191 & 31.760 & & 0.905 & 587.320 
  & 1.970\\
\hline
\hline
\end{tabular}
\end{center}
\end{table}

It is useful to compare theoretical evaluations in such a parameter regime with current experimental constraints imposed on the interaction of top quark with charged Higgs boson.  As soon as $m_{H^\pm}< m_{top}$ for all BP's (see Table \ref{tab:pred}), the main ${H^\pm}$ production mechanism is top-quark decay to $b$ and $H^+$. In the hadronic decay mode $H^+ \to c \bar{s}$ which is dominant for  $\tan \beta<5$ the upper limit on the level of 1--5\% (95\% CL) has been set by the ATLAS experiment on 
$BR(t \to H^+b)$ assuming that $BR(H^+ \to c \bar{s})$=1 at the energy $\sqrt{s}$=7 TeV  \cite{5}. Branhching ratios of the process $t \to b H^+$ are 5.4\% and 2.6\% for BP1 and BP2, correspondingly.
However, for our specific scenario the main decay of charged Higgs boson is to pseudoscalar $A$ and $W^+$ boson (about 90--99\%), and therefore comparison for $\tan \beta<$5 is rather ambiguous.
For $\tan \beta>$5,
the ATLAS and CMS experiments determined an upper limits on $BR(t \to H^+ b)BR (H^+ \to \tau^+ \nu_\tau)$ on the level 1.3--0.2\% \cite{5} and 1.2--0.5\% \cite{4}, correspondingly. Numerical estimations give the value of $BR(t \to H^+ b)BR (H^+ \to \tau^+ \nu_\tau)$ of about 
0.13\% 
for BP3 and 
0.09\% 
for BP4.

\section{\large SUMMARY}
\label{concl}
In conclusion we summarize the results of 28 GeV $\mu^+ \mu^-$ excess identification in the CMS data
as the MSSM CP-odd Higgs boson.
Additive one-loop threshold corrections coming from dimension-six effective operators are taken into account under the assumption that the relevant radiative corrections to the Higgs boson masses come from stop- and sbottom sectors.  
It was found that the MSSM Higgs bosons could be rather light, near the electroweak scale that corresponds to the non-decoupling regime.
The following conclusions can be drawn from a scan of the MSSM parameter space and cross section calculations
\begin{enumerate}
\item[--] light pseudoscalar with the mass $M_A=$28 GeV can be embedded in the two-doublet MSSM Higgs sector extended by dimension-six effective operators respecting the alignment limit for $h$(125 GeV) state in a rather specific range of parameter space, when the superparticle mass scale is around 1--2 TeV, $\tan \beta \sim $ 2--5 and soft SUSY breaking parameters $A_{t,b}, \mu$ are large, from 3 TeV to 9 TeV;
\item[--] such range of the MSSM parameter space is at the limit of fulfillment of the vacuum stability and perturbative unitarity conditions;
\item[--] due to appearance of the decay channel $h \to A A$, the total width $\Gamma_h$ of $h$(125 GeV) state becomes of the order of 1 GeV. Experimental precision on $\Gamma_h$ from on-shell measurements \cite{onshell} of the width corresponds to this value which is, however, worse than the bound from the analyses beyond the infinitely small width approximation \cite{offshell} made under the assumption of the SM coupling structure. Cross section calculations at the tree level for the partonic level signal in $pp \to  \mu^+ \mu^- b \bar{b}$ at the energies $\sqrt{s}=$8 and 13 TeV give signal cross sections by a factor of 2--5 smaller than the experimentally observed cross section of a few fb;
\item[--]
at the same time, numerical estimations based on charged Higgs boson production due to top quark decay are in agreement with current LHC constraints.
\end{enumerate}

\subsection*{Acknowledgments}

The authors are grateful to H. Bahl, O. Kodolova, A. Nikitenko and G. Weiglein for useful discussions. This work was supported by the Russian Science Foundation Grant No. 16-12-10280.


\begin{thebibliography}{0}    

\bibitem{28GeV}
CMS Collaboration, 
{\it Search for resonances in the mass spectrum of muon
pairs produced in association with $b$ quark jets in
proton-proton collisions at $\sqrt{s}$ = 8 and 13 TeV},
J. High
Energy Phys. 11 (2018) 161;
%
{\it Search for a light pseudoscalar Higgs boson produced in association with bottom quarks in pp collisions at $\sqrt{s}$=8 TeV},
J. High Energy Phys. 1711 (2017) 010.

\bibitem{lhc-higgs1}
ATLAS Collaboration, 
{\it Observation of a new particle in the search for the Standard Model Higgs boson with the ATLAS detector at the LHC},
Phys. Lett. B \textbf{716}, 1 (2012). 

\bibitem{lhc-higgs2}
CMS Collaboration, 
{\it Observation of a new boson at a mass of 125 GeV with the CMS experiment at the LHC},
Phys. Lett. B {\bf716}, 30 (2012).

\bibitem{jhep_atlas_cms1}
ATLAS and CMS Collaborations, 
{\it Measurements of the Higgs boson production and decay rates and constraints on its couplings from a combined ATLAS and CMS analysis of the LHC $pp$ collision data at $\sqrt{s}$=7 and 8 TeV}, 
J. High Energy Phys. 1608 (2016) 045.

\bibitem{jhep_atlas_cms2}
ATLAS and CMS Collaborations, 
{\it Combined measurement of the Higgs boson mass in $pp$ collisions at $\sqrt{s}$=7 and 8 TeV with the ATLAS and CMS Experiments},
Phys. Rev. Lett. {\bf 114}, 191803 (2015).

\bibitem{own3}
M.~N. Dubinin and E. Yu. Petrova, 
{\it Scenarios with low mass Higgs bosons in the heavy supersymmetry},
Int. J. Mod. Phys. A {\bf 33}, 1850150 (2018). 

\bibitem{LC_atlas_cms1}
J. Brandstetter, 
{\it Higgs boson results on couplings to fermions, CP parameters and perspectives for HL-LHC} (ATLAS and CMS), 
arXiv:1801.07926v1 [hep-ex].

\bibitem{LC_atlas_cms2}
M. Malberti, 
{\it SM Higgs boson measurements at CMS}, 
Nuovo Cimento C {\bf 40}, 182 (2017).

\bibitem{mssm1}
A. Djouadi, 
{\it The anatomy of electro-weak symmetry breaking. II. The Higgs bosons in the minimal supersymmetric model},
Phys. Rept. {\bf 459}, 1 (2008).

\bibitem{mssm2}
H. Haber and G. Kane,
{\it The search for supersymmetry: probing physics beyond the Standard Model},
Phys. Rept. {\bf 117}, 75 (1985). 


\bibitem{pdg}
C. Patrignany \textit{et al.} (Particle Data Group),
Review of Particle Physics, 
Chin. Phys. C {\bf 40}, 100001 (2016). 


\bibitem{B2017}
M. Misiak and M. Steinhauser,
{\it Weak radiative decays of the $B$ meson and bounds on $M_{H^\pm}$ in the Two-Higgs-Doublet Model},
Eur. Phys. J. C {\bf 77}, 201  (2017).

\bibitem{searchtau19}
CMS Collaboration, {\it Search for a low-mass $\tau^+ \tau^-$ resonance in association with a bottom quark in proton-proton collisions at $\sqrt{s}$=13 TeV},
J. High Energy Phys. 1905 (2019) 210.

\bibitem{searchtau17}
CMS Collaboration,
{\it Search for a light pseudoscalar Higgs boson produced in
association with bottom quarks in $pp$ collisions at
$\sqrt{s}$=8 TeV}, 
J. High Energy Phys. 1711 (2017) 010.

\bibitem{belle19}
Belle Collaboration,	
{\it Search for a light CP-odd Higgs boson and low-mass dark matter at the Belle experiment}, 
Phys. Rev. Lett. \textbf{122}, 011801 (2019).


\bibitem{vysotsky}
S.~I. Godunov, V.~A. Novikov, M.~I. Vysotsky, and E.~V. Zhemchugov,
{\it Dimuon resonance near 28 GeV and the muon anomaly},
JETP Lett. {\bf 109}, 358 (2019). 

\bibitem{kazakov}
C. Beskidt, W. de Boer, and D.I. Kazakov,
{\it Can we discover a light singlet-like NMSSM Higgs boson at the LHC?},
Phys. Lett. B {\bf 782}, 69 (2018). 

\bibitem{benchmark}
M. Carena, S. Heinemeyer, O. Stal, C.~E.~M. Wagner, and G. Weiglein, 
{\it MSSM Higgs boson searches at the LHC: benchmark scenarios after the discovery of a Higgs-like particle},
Eur. Phys. J. C {\bf 73}, 2552 (2013). 

\bibitem{cw73}
S. Coleman and E. Weinberg, 
{\it Radiative corrections as the origin of spontaneous symmetry breaking},
Phys. Rev. D {\bf 7}, 1888 (1973).


\bibitem{haber_hempfling}
H.~E.~Haber and R. Hempfling, 
{\it The Renormalization group improved Higgs sector of the minimal supersymmetric model},
Phys.~Rev. D {\bf 48}, 4280 (1993). 

\bibitem{eff_potential1}
M. Carena, J. Ellis, A. Pilaftsis, and C.~E.~M. Wagner, 
{\it Renormalization group improved effective potential for the MSSM Higgs sector with explicit CP violation},
Nucl. Phys. B {\bf 586}, 92 (2000).

\bibitem{eff_potential2}
S.~Y. Choi, M. Drees, and J.~S. Lee, 
{\it Loop corrections to the neutral Higgs boson sector of the MSSM with explicit CP violation},
Phys. Lett. B {\bf 481}, 57 (2000). 

\bibitem{eff_potential3}
M. Carena, M.~Quiros and C.~E.~M.~Wagner, 
{\it Effective potential methods and the Higgs mass spectrum in the MSSM},
Nucl. Phys. B {\bf 461}, 407 (1996). 

\bibitem{eff_potential4}
M.~Carena, J.~R.~Espinosa, M.~Quiros, and C.~E.~M.~Wagner, 
{\it Analytical expressions for radiatively corrected Higgs masses and couplings in the MSSM},
Phys. Lett. B {\bf 355}, 209 (1995). 

\bibitem{own1}
E. Akhmetzyanova, M. Dolgopolov, and M. Dubinin, 
{\it Higgs bosons in the two-doublet model with CP violation},
Phys. Rev. D {\bf 71}, 075008 (2005). 

\bibitem{own2}
E. Akhmetzyanova, M. Dolgopolov, and M. Dubinin, 
{\it Violation of CP invariance in the two-doublet Higgs sector of the MSSM},
Phys. Part. Nucl. {\bf 37}, 677 (2006).

\bibitem{sim_sc}
M.~N. Dubinin and E.~Yu. Petrova, 
{\it Simplified parametric scenarios of the Minimal Supersymmetric Standard Model after the discovery of the Higgs boson},
Yad. Phys. {\bf 79}, 302 (2016).

\bibitem{dim_six}
M.~N. Dubinin and E. Yu. Petrova, 
{\it Radiative corrections to Higgs boson masses for the MSSM Higgs potential with dimension-six operators},
Phys. Rev. D {\bf 95}, 055021 (2017). 

\bibitem{unitarity_1}
B. W. Lee, C. Quigg, and H. B. Thacker, 
{\it Weak interactions at very high-energies: the role of the Higgs boson mass},
Phys. Rev. D {\bf 16}, 1519 (1977). 

\bibitem{unitarity_21}
I. F. Ginzburg and I. P. Ivanov, 
{\it Tree-level unitarity constraints in the most general 2HDM},
Phys. Rev. D {\bf 72}, 115010 (2005). 

\bibitem{unitarity_22}
A. G. Akeroyd, A. Arhrib, and E. Naimi,
{\it Note on tree level unitarity in the general two Higgs doublet model},
Phys. Lett. B {\bf 490}, 119 (2000). 

\bibitem{staub}
F. Staub,
{\it Theoretical constrants on supersymmetric models: perturbative unitarity vs. vacuum stability},
Phys. Lett. B {\bf 789}, 2013 (2019).

\bibitem{kanemura}
S. Kanemura and K. Yagyu,
{\it Unitarity bound in the most general two-Higgs doublet model},
Phys. Lett. B {\bf 751}, 289 (2015).

\bibitem{vacuum}
W.~G. Hollik, G. Weiglein, and J. Wittbrodt, 
{\it Impact of vacuum stability constraints on the
phenomenology of supersymmetric models},
J. High Energy Phys. 1903 (2019) 109. 

\bibitem{own-min}
M.~N. Dubinin and E.~Yu. Petrova, 
{\it Vacuum stability with the effective six-pointed couplings of the Higgs bosons in the heavy supersymmetry},
EPJ Web Conf. {\bf 158}, 02005 (2017).


\bibitem{alignment1}
M.~Carena, H.~E.~Haber, I.~Low, N.~R. Shah, and C.~E.~M. Wagner,
{\it Complementarity between nonstandard Higgs boson searches and precision Higgs boson measurements in the MSSM},
Phys. Rev. D {\bf 91}, 035003 (2015). 

\bibitem{alignment2}
D.~Asner {\it et al.}, 
ILC Higgs white paper,
in {\it 
2013 Community Summer Study on the Future of U.S. Particle Physics: Snowmass on the Mississippi (CSS2013), Minneapolis 2013},
edited by N.~A. Graf,
M.~E. Peskin, and J.~L. Rosner (The Division of Particles and Fields of the American Physical Society, Minneapolis, 2013), p. 1.

\bibitem{comphep}
CompHEP Collaboration, 
{\it CompHEP 4.4: Automatic computations from Lagrangians to events}, Nucl. Instrum. Methods \textbf{A534}, 250 (2004); 
A. Pukhov \textit{et al.}, {\it CompHEP -- a package for evaluation of Feynman diagrams and integration over multi-particle phase space. 
    User's manual for version 3.3}, arXiv:hep-ph/9908288.
    
\bibitem{5}
ATLAS Collaboration, 
{\it Search for charged Higgs bosons decaying via $H^\pm \to \tau^\pm \nu$ in fully hadronic final states using $pp$ collision data at $\sqrt{s}$=8 TeV with the ATLAS detector},
J. High Energy Phys. 03 (2015) 088; 
{\it Search for a light charged Higgs boson in the decay channel $H^+ \to c \bar{s}$ in $t \bar{t}$ events using pp collisions at $\sqrt{s}$ = 7 TeV with the ATLAS detector},
Eur. Phys. J. C \textbf{73}, 2465 (2013).

\bibitem{4}
CMS Collaboration, 
{\it Search for a charged Higgs boson in $pp$ collisions at $\sqrt{s}$=8 TeV},
J. High Energy Phys. 1511 (2015) 018.

\bibitem{onshell}
CMS Collaboration,
{\it Measurements of properties of the Higgs boson decaying into the four-lepton final state in pp collisions at $\sqrt{s}$=13 TeV},
J. High Energy Phys. 1711 (2017) 047;
%
ATLAS Collaboration,
{\it Measurement of the Higgs boson mass from the $H \to \gamma \gamma$ and $H \to ZZ^* \to 4l$ channels with the ATLAS detector using 25 fb$^{-1}$ of $pp$ collision data},
Phys. Rev. D {\bf 90} 052004 (2014).

\bibitem{offshell}
J.~M. Campbell, R.~K. Ellis, and C. Williams, 
{\it Bounding the Higgs width at the LHC using full analytic results for $gg \to e^- e^+ \mu^- \mu^+$},
J. High Energy Phys. 04 (2014) 060;
F. Caola and K. Melnikov, 
{\it Constraining the Higgs boson width with $ZZ$ production at the LHC},
Phys. Rev. D {\bf 88}, 054024 (2013);
N. Kauer and G. Passarino, 
{\it Inadequacy of zero-width approximation for a light Higgs boson signal},
J. High Energy Phys. 08 (2012) 116.


\end{thebibliography}
\end{document}